\documentclass{article}
\usepackage[utf8]{inputenc}
\usepackage{amsmath,amssymb,array}
\RequirePackage{fancyvrb}
\RequirePackage{alltt}
\DefineVerbatimEnvironment{example}{Verbatim}{}
\renewenvironment{example*}{\begin{alltt}}{\end{alltt}}
\usepackage{graphicx}
\usepackage{authblk}
\usepackage[verbose=true,letterpaper]{geometry}
\AtBeginDocument{
  \newgeometry{
    textheight=9in,
    textwidth=5.5in,
    top=1in,
    headheight=12pt,
    headsep=25pt,
    footskip=30pt
  }
}
\begin{document}
\title{Approximating the Sum of Independent Non-Identical Binomial Random Variables}

\author[1,2,3]{Boxiang Liu}
\author[4]{Thomas Quertermous}
\affil[ ]{$^1$Departments of Biology, $^2$Pathology, $^3$Genetics, $^4$Medicine - Cardiovascular Medicine, Stanford University}
\affil[ ]{\textit{\{bliu2,tomq1\}@stanford.edu}}

\maketitle

\begin{abstract}
\small The distribution of sum of independent non-identical binomial random variables is frequently encountered in areas such as genomics, healthcare, and operations research. Analytical solutions to the density and distribution are usually cumbersome to find and difficult to compute. Several methods have been developed to approximate the distribution, and among these is the saddlepoint approximation. However, implementation of the saddlepoint approximation is non-trivial and, to our knowledge, an R package is still lacking. In this paper, we implemented the saddlepoint approximation in the \textbf{sinib} package. We provide two examples to illustrate its usage. One example uses simulated data while the other uses real-world healthcare data. The \textbf{sinib} package addresses the gap between the theory and the implementation of approximating the sum of independent non-identical binomials.
\end{abstract}

\section{Introduction}

Convolution of independent non-identical binomial random variables appears in a variety of applications, such as analysis of variant-region overlap in genomics \cite{Schmidt:2015ca}, calculation of bundle compliance statistics in healthcare organizations \cite{Benneyan:2010ex}, and reliability analysis in operations research \cite{Anonymous:iz}. 

Computating the exact distribution of the sum of non-identical independent binomial random variables requires enumeration of all possible combinations of binomial outcomes that satisfy the totality constraint. However, analytical solutions are often difficult to find for sums of greater than two binomial random variables. Several studies have proposed approximate solutions  \cite{Johnson:2005hg,Jolayemi:1992jz}. In particular, Eisinga et al. examined  the saddlepoint approximation, and compared them to exact solutions \cite{Eisinga:2013da}. They note that, in practice, these approximations are often as good as the exact solution and are easy to implement in most statistical software. 

Despite the theoretical development of aforementioned approximate solutions, a software implementation in R is still lacking. The \textbf{stats} package includes functions for the most frequently used distribution such as \textbf{dbinom} and \textbf{dnorm}. In addition, it also include less frequent distributions such as \textbf{pbirthday}. However, it does not contain functions for distribution of sum of independent non-identical binomials. Another package, \textbf{EQL}, provides a \textbf{saddlepoint} function to approximate the mean of i.i.d random variables. However, this function does not apply to the case where the random variables are not identical. In this paper, we address this deficiency by implementing a saddlepoint approximation in the open source package \textbf{sinib} (sum of independent non-identical binomial random variables). The package provides the standard suite of probability (\textbf{psinib}, distribution (\textbf{dsinib}, quantile (\textbf{qsinib}, and random number generator (\textbf{rsinib} functions. The package is accompanied by a detailed documentation, and can be easily integrated into existing applications.

The remainder of this paper is organized as follows, section 2 formulates the distribution of sum of independent non-identical binomial random variables. Section 3 gives an overview of the saddlepoint approximation. Section 4 describes the design and implementation of the \textbf{sinib} package. Section 5 uses two examples to illustrate its usage. As a bonus, we assess the accuracty of saddlepoint approximation in both examples. Section 6 draws final conclusion and discusses possible future development of the package. 

\section{Overview of the distribution}

Suppose $X_1$,...,$X_m$ are independent non-identical binomial random variables, and $S_m = \sum_{i=1}^{m} X_i$. We are interested in finding the distribution of $S_m$. 

\begin{equation}
P(S_m = s) = P(X_1+X_2+...+X_m = s)
\end{equation}

In the special case of $m = 2$, the probability simplifies to 

\begin{equation}
P(S_2=s) = P(X_1+X_2=s) = \sum_{i=0}^s P(X_1=i) P(X_2=s-i)
\label{eq:2}
\end{equation}

Computation of the exact distribution often involves enumerating all possible combinations of each variable that sums to a given value, which becomes infeasible when n is large. A fast recursion method to compute the exact distribution has been proposed \cite{Butler:2016cj,ArthurWoodward:1997en}. The algorithm is as follows: 

\begin{enumerate}
	\item Compute the exact distribution of each $X_i$.
	\item Calculate the distribution of $S_2=X_1+X_2$ using equation \ref{eq:2} and cache the result. 
	\item Calculate $S_r = S_{r-1} + X_i$ for $r = 3,4,...,m$.
\end{enumerate}

Although the recursion speeds up the calculation, studies have shown that the result may be numerically unstable due to round-off error in computing $P(S_r=0)$ if r is large\cite{Yili:al1vGrvv,Eisinga:2013da}. Therefore, approximation methods are still widely used in literature.  

\section{Saddlepoint approximation}

The saddlepoint approximation, first proposed by \cite{Daniels:1954hy} and later extended by \cite{Lugannani:1980jm}, provides highly accurate approximations for the probability and density of any distribution. In brief, let $M(u)$ be the moment generating function, and $K(u) = \log(M(u))$ be the cumulant generating function. The saddlepoint approximation to the PDF of the distribution is given as: 

\begin{equation}
\hat{P}_1(S=s)=\frac{\exp(K(\hat{u}) - \hat{u} s)}{\sqrt{2 \pi K''(\hat{u})}}
\end{equation}

where $\hat{u}$ is the unique value that satisfies $K'(\hat{u})=s$.

\cite{Eisinga:2013da} applied the saddlepoint approximation to sum of independent non-identical binomial random variables. Suppose that $X_i \sim Binomial(n_i, p_i) \text{ for } i = 1,2,...,m$. The cumulant generating function of $S_m = \sum_{i=1}^m X_i$ is:

\begin{equation}
K(u) = \sum_{i=1}^m n_i \ln (1-p_i + p_i \exp(u))
\end{equation}

The first- and second-order derivative of $K(u)$ are:

\begin{equation}
K'(u) = \sum_{i=1}^m n_i q_i
\end{equation}

\begin{equation}
K''(u) = \sum_{i=1}^m n_i q_i (1-q_i)
\end{equation}

where $q_i = p_i exp(u) (1-p_i + p_i \exp(u))$. 

The saddlepoint of $\hat{u}$ can be obtained by solving $K'(\hat{u})=s$. A unique root can always be found because $K(u)$ is strictly convex and therefore $K'(u)$ is monotonically increasing on the real line. 

The above shows the first-order approximation of the distribution. The approximation can be improved by adding a second-order correction term \cite{Anonymous:0hq1uBaf}. 

\begin{equation}
\hat{P}_2(S=s)=\hat{P}_1(S=s)\Big\{1 + \frac{K''''(\hat{u})}{8[K''(\hat{u})]^2} - \frac{5[K'''(\hat{u})]^2}{24[K''(\hat{u})]^3}\Big\}
\end{equation}

where 

$$K'''(\hat{u}) = \sum_{i=1}^m n_i q_i (1-q_i) (1-2q_i)$$

and 

$$K''''(\hat{u}) = \sum_{i=1}^m n_i q_i (1-q_i) [1-6q_i(1-q_i)]$$

Although the saddlepoint equation cannot be solved at boundaries $s = 0$ and $s=\sum_{i=1}^m n_i$, their exact probabilities can be computed easily: 

\begin{equation}
P(S=0) = \prod_{i=1}^m (1-p_i)^{n_i}
\end{equation}

\begin{equation}
P(S=\sum_{i=1}^m n_i)=\prod_{i=1}^m p_i^{n_i}
\end{equation}

Incorporation of boundary solutions into the approximation gives: 

\begin{equation}
\bar{P}(S=s)=
\begin{cases}
P(S=0), & s=0 \\
[1-P(S=0)-P(S=\sum_{i=1}^m n_i)] \frac{\hat{P}_2(S=s)}{\sum_{i=1}^{\sum_{i=1}^m n_i-1} \hat{P}_2(S=i)}, & 0 < s < \sum_{i=1}^m n_i \\
P(S=\sum_{i=1}^m n_i), & s=\sum_{i=1}^m n_i 
\end{cases}
\label{eq:10}
\end{equation}

We implemented equation \ref{eq:10} as the final approxmation of the probability density function. For the cumulative density, \cite{Anonymous:0hq1uBaf} gave the following approximator: 

\begin{equation}
\hat{P}_3(S \geq s)=
\begin{cases}
1-\Phi(\hat{w})-\phi(\hat{w}) (\frac{1}{\hat{w}} - \frac{1}{\hat{u_1}}), & \text{if } s \neq E(S) \text{ and } \hat{u} \neq 0 \\
\frac{1}{2} - \frac{1}{\sqrt{2 \pi}} \big[\frac{K'''(0)}{6 K''(0)^{3/2}} - \frac{1}{2 \sqrt{K''(0)}} \big], & \text{otherwise}
\end{cases}
\end{equation}

where $\hat{w}= \text{sign}(\hat{u}) [2 \hat{u} K'(\hat{u}) - 2K(\hat{u})]^{1/2}$ and $\hat{u}_1=[1-\exp(-\hat{u})][K''(\hat{u})]^{1/2}$. The letters $\Phi$ and $\phi$ denotes the probability and density of the standard normal distribution. 

The accuracy can be improved by adding a second-order continuity correction:

\begin{equation}
\hat{P}_4(S \geq s)=\hat{P}_3(S \geq s) - \phi(\hat{w}) \Big [ \frac{1}{\hat{u}_2} \Big ( \frac{\hat{\kappa}_4}{8} - \frac{5 \hat{\kappa}^2_3}{24} \Big ) - \frac{1}{\hat{u}_2^3} - \frac{\hat{\kappa}_3}{2 \hat{u}_2^2} + \frac{1}{\hat{w}^3} \Big]
\label{eq:12}
\end{equation}

where $\hat{u}_2=\hat{u}[K''(\hat{u})]^{1/2}$, $\hat{\kappa}_3=K'''(\hat{u}) [K''(\hat{u})]^{-3/2}$, and $\hat{\kappa}_4 = K''''(\hat{u}) [K''(\hat{u})]^{-2}$.

We implemented equation \ref{eq:12} in the package to approximate the cumulative distribution. 

\section{The \textbf{sinib} package}

The package used only functions in base R and the stats package to minimize compatibility issues. The arguments for the functions in the \textbf{sinib} package are designed to have similar meaning to those in the \textbf{stats} package, thereby minimizing the learning required. To illustrate, we compare the arguments of the \textbf{*binom} and the \textbf{*sinib} functions. 

From the help page of the binomial distribution: 

\begin{itemize}
\item x, q: vector of quantiles.
\item p: vector of probabilities.
\item n: number of observations. 
\item size: number of trials. 
\item prob: probability of success on each trial.
\item log, log.p: logical; if TRUE, probabilities p are given as log(p). 
\item lower.tail: logical; if TRUE (default), probabilities are $P[X \leq x]$, otherwise, $P[X > x]$. 
\end{itemize}

Since the distribution of sum of independent non-identical binomials is defined by a vector of trial and probability pairs (each pair for one constituent binomial), it was neccessary to redefine these arguments in the \textbf{*sinib} functions. Therefore, the following two arguments need to be redefined:

\begin{itemize}
\item size: integer vector of number of trials. 
\item prob: numeric vector of success probabilities.
\end{itemize}

All other arguments remain the same. It is worth noting that when size and prob arguments are given as vectors of length 1, the \textbf{*sinib} functions redues to \textbf{*binom} functions: 

\begin{example}
# Binomial:
dbinom(x = 1, size = 2, prob = 0.5)
[1] 0.5


# Sum of binomials:
library(sinib)
dsinib(x = 1, size = 2, prob = 0.5)
[1] 0.5
\end{example}

With that in mind, the next section shows a few examples to illustrate the usage of \textbf{sinib}. 

\section{Two examples}
\subsection{Sum of two binomials}
We use two examples to illustrate the use of this package, starting from the simplest case of two binomial random variables with the same mean but different sizes, $X \sim Bin(n,p)$ and $Y \sim Bin(m,p)$. The distribution of $S = X+Y$ has an analytical solution, $S \sim Bin(m+n,p)$. We can therefore use different combinations of $(m,n,p)$ to assess the accuracy of the saddlepoint approximation of the CDF. We use $m,n = \{10, 100, 1000\}$ and $p = \{0.1, 0.5, 0.9\}$ to assess the approximation. The ranges of m and n are chosen to be large and the value of p are chosen to represent both boundaries. 

\begin{example}
library(foreach)
library(data.table)
library(cowplot)
library(sinib)

# Comparison of CDF between truth and approximation:
data=foreach(m=c(10,100,1000),.combine='rbind')%do%{
	foreach(n=c(10,100,1000),.combine='rbind')%do%{
		foreach(p=c(0.1, 0.5, 0.9),.combine='rbind')%do%{
			a=pbinom(q=0:(m+n),size=(m+n),prob = p)
			b=psinib(q=0:(m+n),size=as.integer(c(m,n)),prob=c(p,p))
			data.table(s=seq_along(a),truth=a,approx=b,m=m,n=n,p=p)
		}
	}
}

ggplot(data,aes(x=truth,y=approx,color=as.character(p)))+
	geom_point(alpha=0.5)+
	facet_grid(m~n)+
	theme_bw()+
	scale_color_discrete(name='prob')+
	xlab('Truth')+ylab('Approximation')
\end{example}

\begin{figure}[h]
\includegraphics[width=\textwidth]{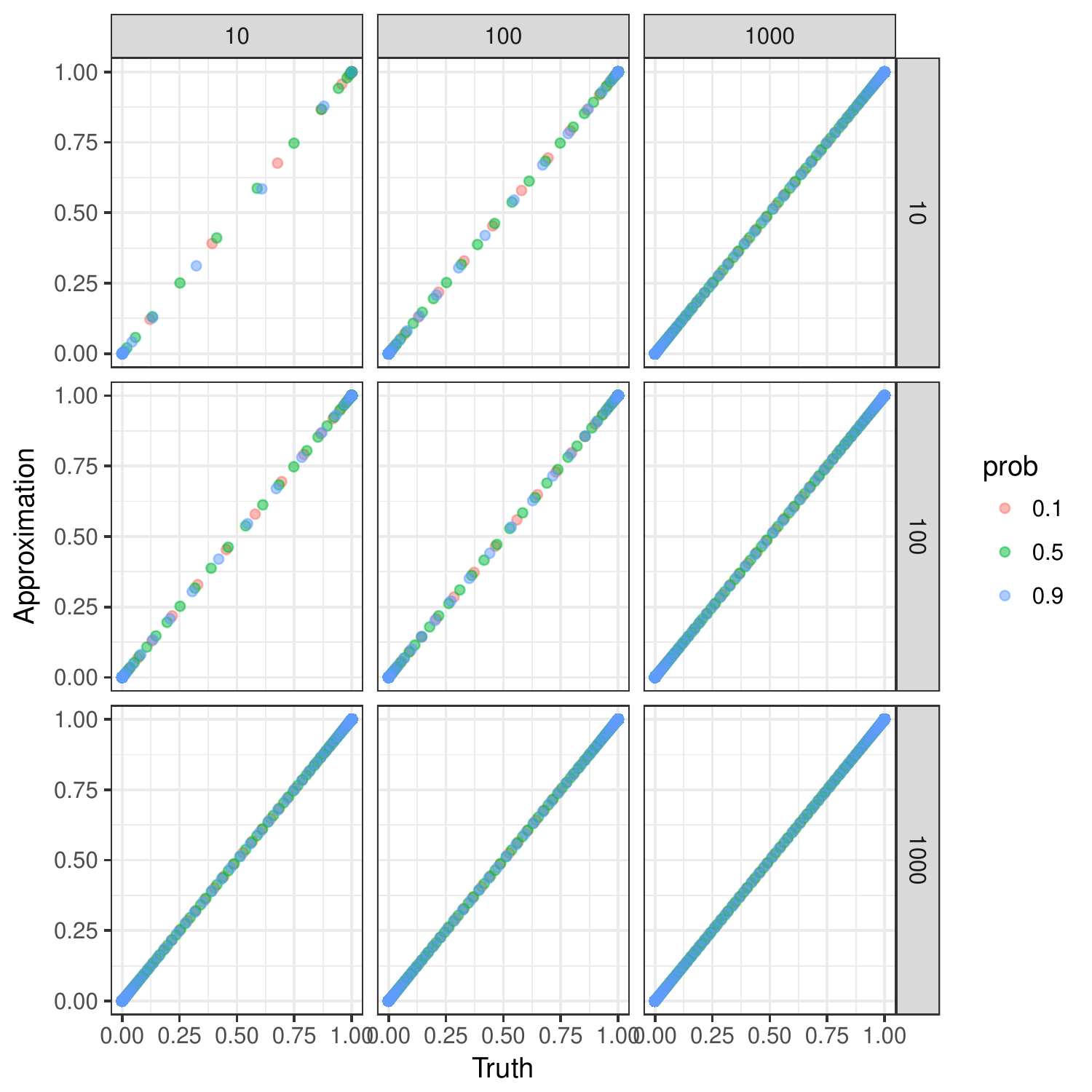}
\caption{Comparison of CDF between truth and approximation}
\label{fig:1}
\end{figure}

Figure \ref{fig:1} shows that the approximations are close to the ground truths across a range of parameters. We can further examine the accuracy by looking at the differences between the approximations and the ground truths. 

\begin{example}
p2=ggplot(data[m==100&n==100],
	aes(x=s,y=truth-approx,color=as.character(p)))+
	geom_point(alpha=0.5)+theme_bw()+
	scale_color_discrete(name='prob')+
	xlab('Quantile')+ylab('Truth-Approximation')+
	geom_vline(xintercept=200*0.5,color='green',linetype='longdash')+
	geom_vline(xintercept=200*0.1,color='red',linetype='longdash')+
	geom_vline(xintercept=200*0.9,color='blue',linetype='longdash')
\end{example}

\begin{figure}[h]
\includegraphics[width=\textwidth]{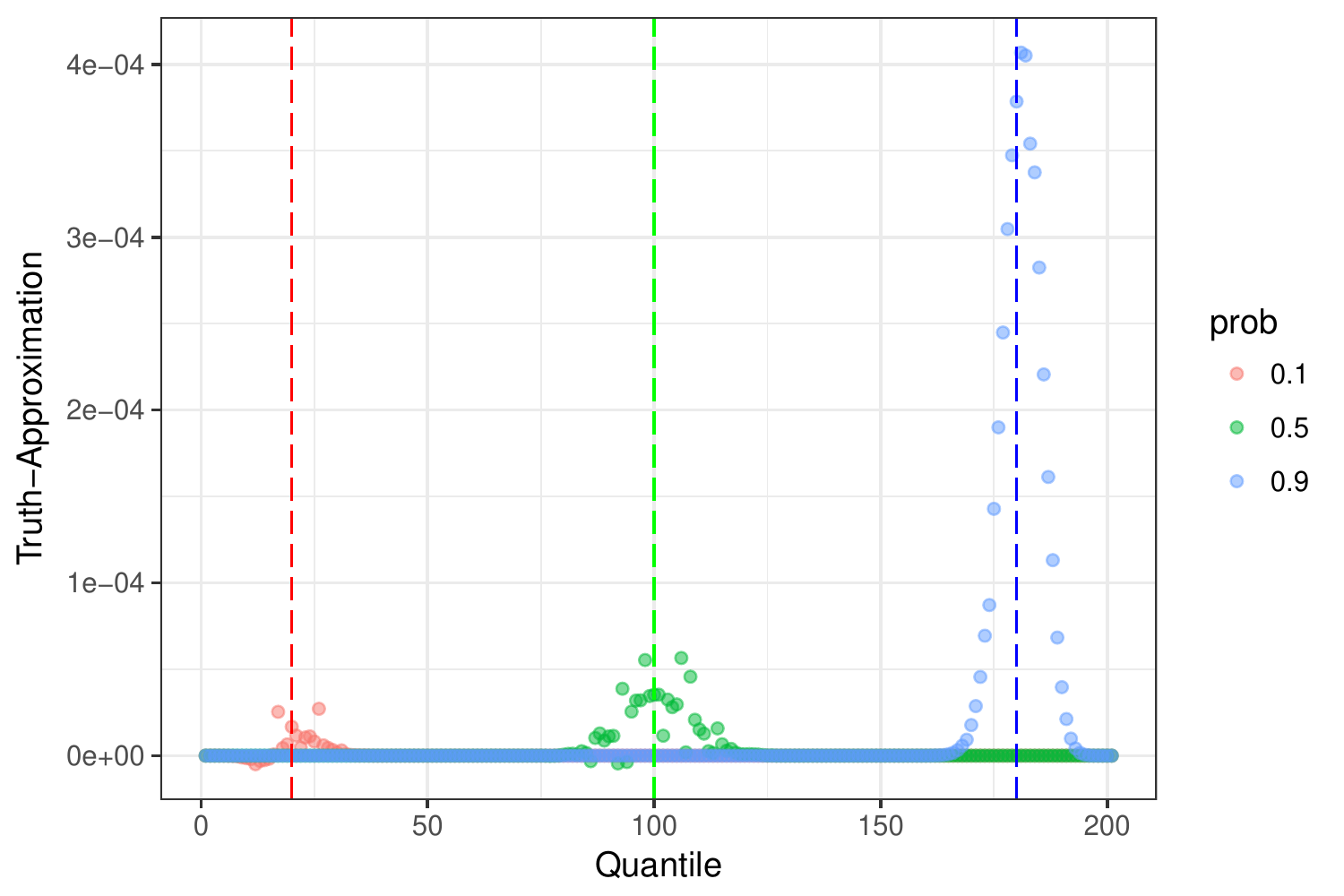}
\caption{Difference in CDF between the ground truth and the approximation}
\label{fig:2}
\end{figure}

Figure \ref{fig:2} shows the difference between the truth and the approximation for $m=n=100$. The dashed line indicate the mean of each random variable. The approximations perform well overall. The largest difference occurs around the mean (dashed lines), but the largest deviation is less than 5e-4. We can also examine the approximation for the PDF. 

\begin{example}
# Comparison of PDF between truth and approximation:
data=foreach(m=c(10,100,1000),.combine='rbind')%do%{
	foreach(n=c(10,100,1000),.combine='rbind')%do%{
		foreach(p=c(0.1, 0.5, 0.9),.combine='rbind')%do%{
			a=dbinom(x=0:(m+n),size=(m+n),prob = p)
			b=dsinib(x=0:(m+n),size=as.integer(c(m,n)),prob=c(p,p))
			data.table(s=seq_along(a),truth=a,approx=b,m=m,n=n,p=p)
		}
	}
}

ggplot(data,aes(x=truth,y=approx,color=as.character(p)))+
	geom_point(alpha=0.5)+facet_grid(m~n)+
	theme_bw()+scale_color_discrete(name='prob')+
	xlab('Truth')+ylab('Approximation')
\end{example}

\begin{figure}[h]
\includegraphics[width=\textwidth]{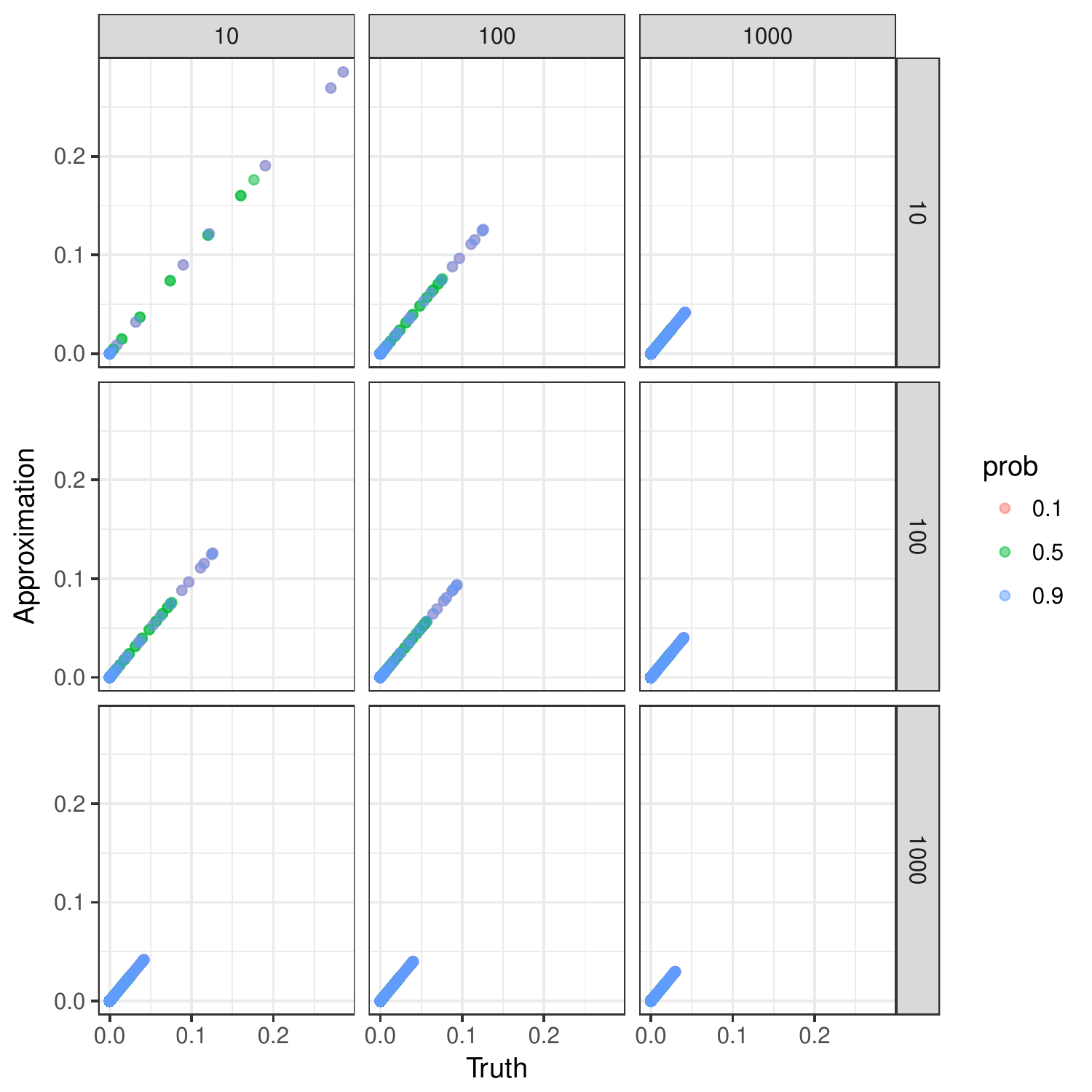}
\caption{Comparison of PDF between truth and approximation}
\label{fig:3}
\end{figure}

\begin{figure}[h]
\includegraphics[width=\textwidth]{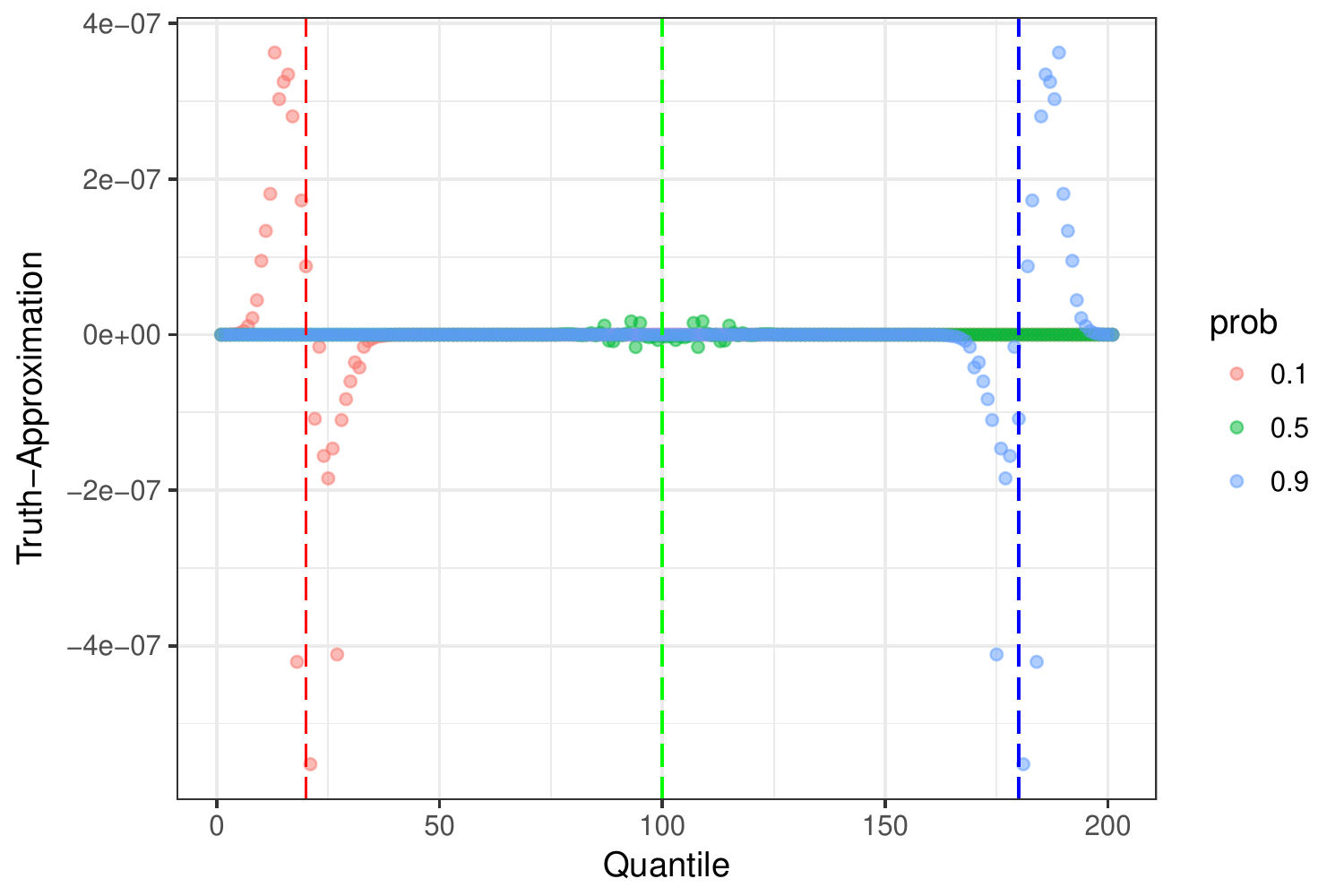}
\caption{Difference in PDF between truth and approximation}
\label{fig:4}
\end{figure}

Figure \ref{fig:3} shows that the approximations and the ground truths are similar. Once again, we examine the difference between the truth and the approximation. One example for $m=n=100$ is shown in figure \ref{fig:4}. As before, the approximation degrades around the mean, but the largest deviation is less than 4e-7.

\begin{example}
p4=ggplot(data[m==100&n==100],
	aes(x=s,y=truth-approx,color=as.character(p)))+
	geom_point(alpha=0.5)+theme_bw()+
	scale_color_discrete(name='prob')+
	xlab('Quantile')+ylab('Truth-Approximation')+
	geom_vline(xintercept=200*0.5,color='green',linetype='longdash')+
	geom_vline(xintercept=200*0.1,color='red',linetype='longdash')+
	geom_vline(xintercept=200*0.9,color='blue',linetype='longdash')
\end{example}

\subsection{Healthcare monitoring}
In the second example, we used a health system monitoring dataset by \cite{Benneyan:2010ex}. Suppose $n_i$ and $p_i$ take the following values. 

\begin{example}
size=as.integer(c(12, 14, 4, 2, 20, 17, 11, 1, 8, 11))
prob=c(0.074, 0.039, 0.095, 0.039, 0.053, 0.043, 0.067, 0.018, 0.099, 0.045)
\end{example}

Since it is difficult to find an analytical solution to the density, we estimated the density with simulation (1e8 trials) and treated it as the ground truth. We then compared simulations with 1e3, 1e4, 1e5, and 1e6 trials, as well as the saddlepoint approximation, to the ground truth. 

\begin{example}
# Sinib:
approx=dsinib(0:sum(size),size,prob)
approx=data.frame(s=0:sum(size),pdf=approx,type='saddlepoint')

# Simulation:
data=foreach(n_sim=10^c(3:6,8),.combine='rbind')%do%{
	n_binom=length(prob)
	set.seed(42)
	mat=matrix(rbinom(n_sim*n_binom,size,prob),nrow=n_binom,ncol=n_sim)
	
	S=colSums(mat)
	sim=sapply(X = 0:sum(size), FUN = function(x) {sum(S==x)/length(S)})
	data.table(s=0:sum(size),pdf=sim,type=n_sim)
}

data=rbind(data,approx)
truth=data[type=='1e+08',]

merged=merge(truth[,list(s,pdf)],data,by='s',suffixes=c('_truth','_approx'))
merged=merged[type!='1e+08',]

ggplot(merged,aes(pdf_truth,pdf_approx))+
	geom_point()+
	facet_grid(~type)+
	geom_abline(intercept=0,slope=1)+
	theme_bw()+
	xlab('Truth')+ylab('Approx')
\end{example}

\begin{figure}[h]
\includegraphics[width=\textwidth]{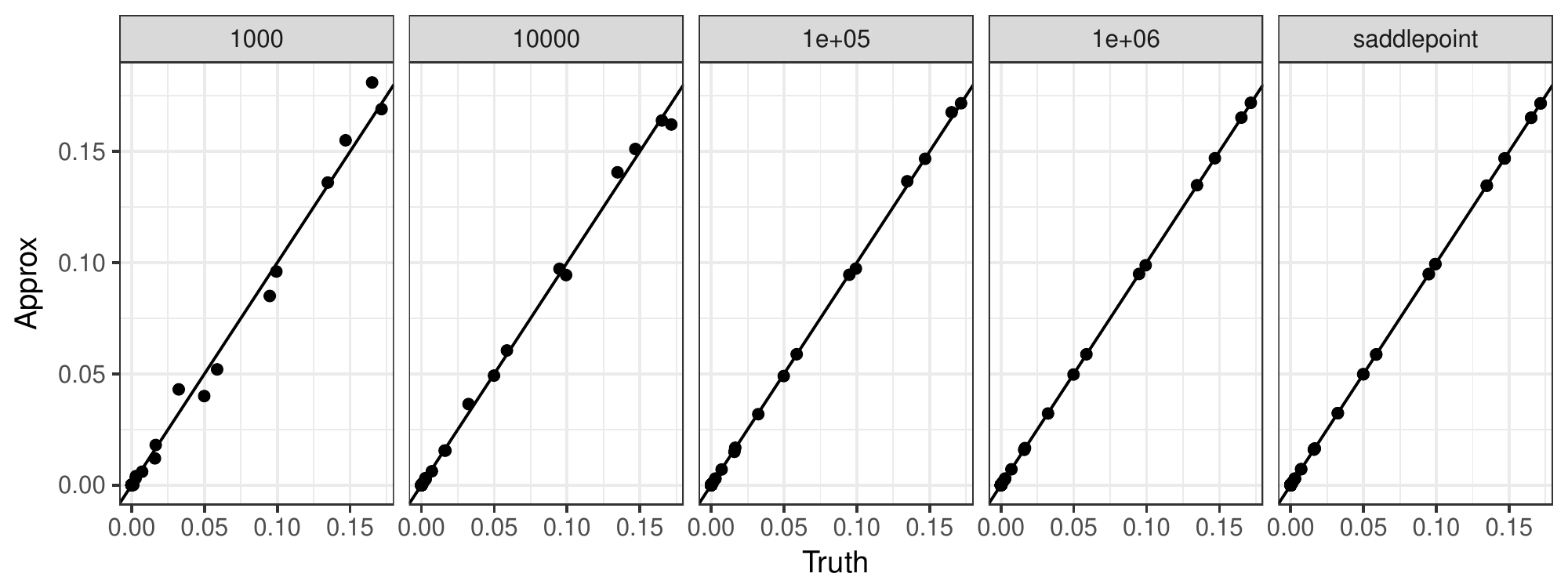}
\caption{Comparison of PDF between truth and approximation}
\label{fig:5}
\end{figure}

Figure \ref{fig:5} shows that the simulation with 1e6 trials and the saddlepoint approximation are visually indistinguishable from the ground truth, while simulations with smaller sizes shows clear deviations from the truth. To be precise, we plotted the difference in PDF between the truth and the approximation.

\begin{example}
merged[,diff:=pdf_truth-pdf_approx]

ggplot(merged,aes(s,diff))+
	geom_point()+
	facet_grid(~type)+
	theme_bw()+
	xlab('Quantile')+ylab('Truth-Approx')
\end{example}

\begin{figure}[h]
\includegraphics[width=\textwidth]{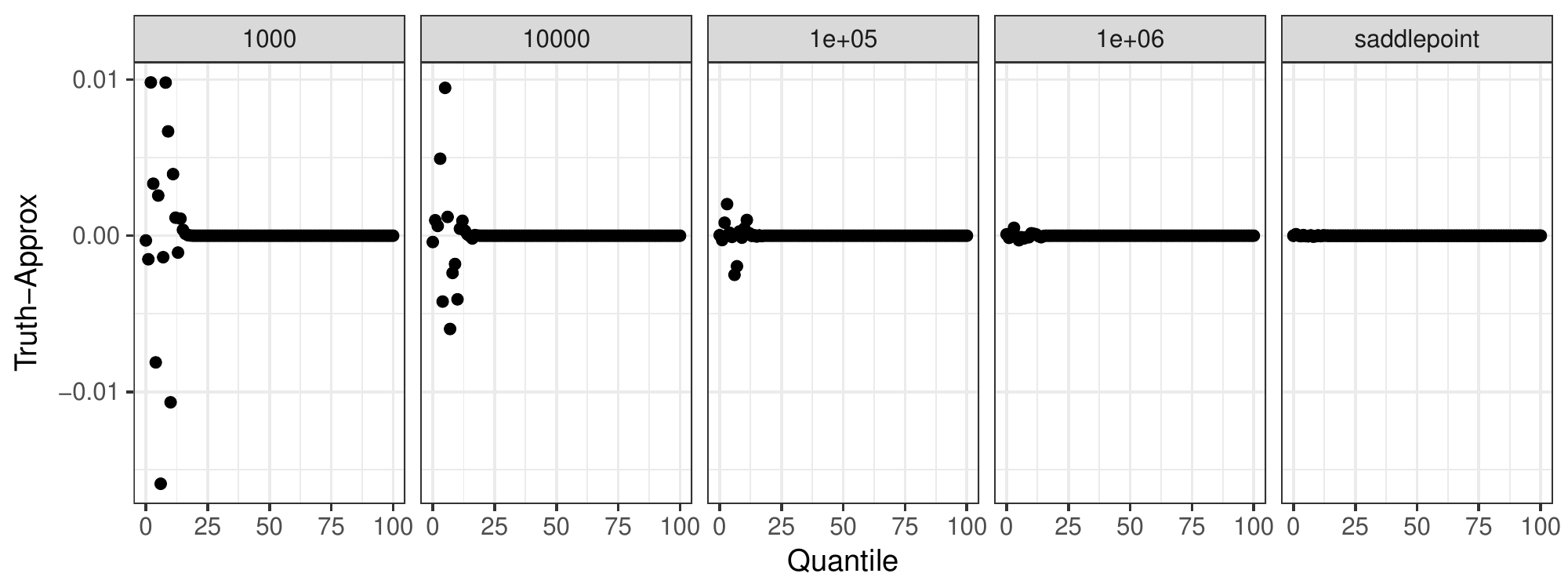}
\caption{Comparison of PDF between truth and approximation}
\label{fig:6}
\end{figure}

Figure \ref{fig:6} shows that that the saddlepoint method and the simulation with 1e6 both provide good approximations, while simulations of smaller sizes show clear deviations. Note that the saddlepoint approximation is 5 times faster than the simulation of 1e6 trials.

\begin{example}
ptm=proc.time()
n_binom=length(prob)
mat=matrix(rbinom(n_sim*n_binom,size,prob),nrow=n_binom,ncol=n_sim)
S=colSums(mat)
sim=sapply(X = 0:sum(size), FUN = function(x) {sum(S==x)/length(S)})
proc.time()-ptm
#    user  system elapsed 
#   1.008   0.153   1.173 
 
ptm=proc.time()
approx=dsinib(0:sum(size),size,prob)
proc.time()-ptm
#   user  system elapsed 
#  0.025   0.215   0.239
\end{example}

\section{Conclusion and future directions}

In this paper, we presented an implementation of the saddlepoint method to approximate the distribution of sum of independent and non-identical binomials. We assessed the accuracy of the method by, first, comparing it with the analytical solution on a simple case of two binomials, and second, with the simulated ground truth on a real-world dataset in healthcare monitoring. These assessments suggest that, while the saddlepoint approximation deviates from the ground truth around the means, it generally provides an approximation superior to simulation in terms of both speed and accuracy. Overall, the \textbf{sinib} package addresses the gap between the theory and implementation on the approximation of sum of independent and non-identical binomial random variables. 

In the future, we aim to explore other approximation methods such as the Kolmogorov approximation and the Pearson curve approximation described by \cite{Butler:2016cj}.
\bibliographystyle{unsrt}
\bibliography{RJreferences}
\end{document}